\documentclass{iaus}
\usepackage{graphicx}

\title[GAMA] 
{Galaxy And Mass Assembly (GAMA)}

\author[Simon P. Driver and the GAMA team]   
{Simon P. Driver$^1$
 \and the GAMA team\thanks{
I.K.Baldry (LJMU),
S.Bamford(Nott),
J.Bland-Hawthorn(USyd),
T.Bridges(AAO),
E.Cameron(StA),
C.Conselice(Nott),
W.J.Couch(Swinburne),
S.Croom(USyd),
N.J.G.Cross(Edin),
S.P.Driver (StA),
L.Dunne(Nott.),
S.Eales(Cardiff),
E.Edmondson(Ports),
S.C.Ellis(USyd),
C.S.Frenk(Durham),
A.W.Graham(Swinburne),
H.Jones(AAO),
D.Hill(StA),
A.Hopkins(USyd),
E.van Kampen(Inns),
K.Kuijken(Leiden),
O.Lahav(UCL),
J.Liske(ESO),
J.Loveday(Sussex),
B.Nichol(Ports.),
P.Norberg(Edin),
S.Oliver(Sussex),
H.Parkinson(Edin),
J.A.Peacock(Edin),
S.Phillipps(Bristol),
C.C.Popescu(UCLan),
M.Prescott(LJMU),
R.Proctor(Swinburne),
R.Sharp(AAO),
L.Staveley-Smith(UWA),
W.Sutherland(QMW),
R.J.Tuffs(MPIK),
S.Warren(Imperial).
}}

\affiliation{$^1$(SUPA) School of Physics and Astronomy, University of St Andrews\\ St Andrews, Fife, KY16 8RS, Scotland \\email: {\tt spd3@st-and.ac.uk}}

\pubyear{2008}
\volume{254}  
\pagerange{119--126}
\jname{The Galaxy Disc in Cosmological Context}
\editors{J. Anderson, J. Bland-Hawthorn \& B. Nordstrom, eds.}
\begin{document}

\maketitle

\begin{abstract}
The GAMA survey aims to deliver 250,000 optical spectra
(3--7\AA~resolution) over 250 sq.\ degrees to spectroscopic limits of
$r_{AB} <19.8$ and $K_{AB}<17.0$ mag. Complementary imaging will be
provided by GALEX, VST, UKIRT, VISTA, HERSCHEL and ASKAP to comparable
flux levels leading to a definitive multi-wavelength galaxy
database. The data will be used to study all aspects of cosmic
structures on 1kpc to 1Mpc scales spanning all environments and out to
a redshift limit of $z \approx 0.4$. Key science drivers include the
measurement of: the halo mass function via group velocity dispersions;
the stellar, HI, and baryonic mass functions; galaxy component
mass-size relations; the recent merger and star-formation rates by
mass, types and environment. Detailed modeling of the spectra, broad
SEDs, and spatial distributions should provide individual star
formation histories, ages, bulge-disc decompositions and stellar
bulge, stellar disc, dust disc, neutral HI gas and total dynamical
masses for a significant subset of the sample ($\sim 100$k) spanning
both the giant and dwarf galaxy populations. The survey commenced
March 2008 with 50k spectra obtained in 21 clear nights using the
Anglo Australian Observatory's new multi-fibre-fed bench-mounted
dual-beam spectroscopic system (AA$\Omega$).

\end{abstract}

\firstsection 

\begin{figure}[t]
\begin{center}
 \includegraphics[width=5.0in]{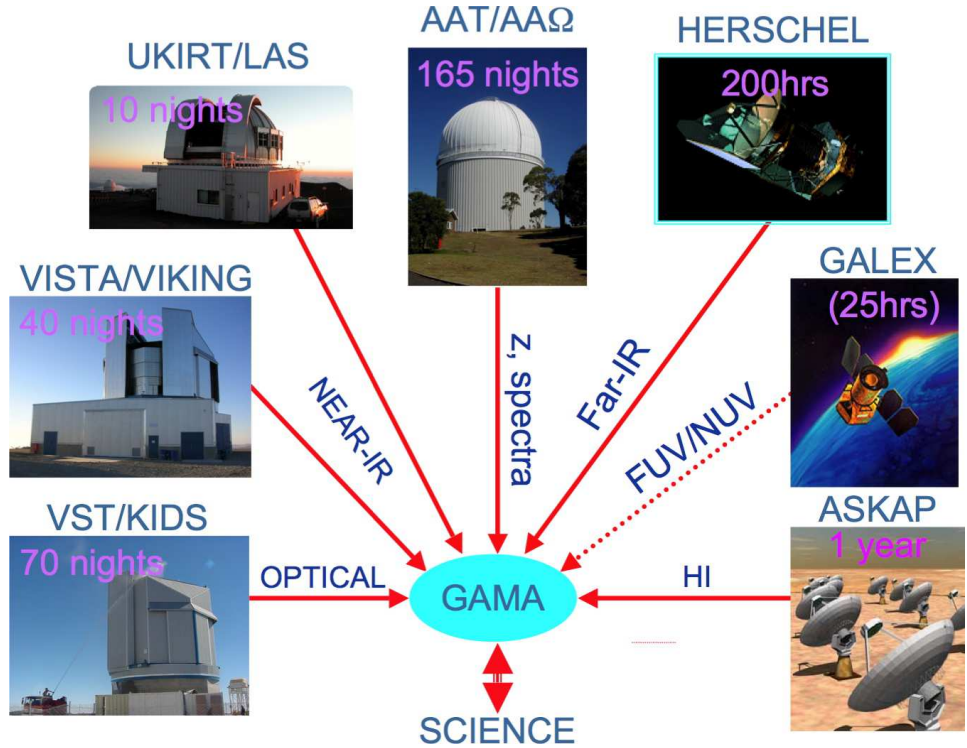} 
 \caption{Facilities contributing to the final GAMA database.}
   \label{fig1}
\end{center}
\end{figure}

\begin{figure}[t]
\vspace*{-4.0 cm}
\begin{center}
 \includegraphics[width=5.5in]{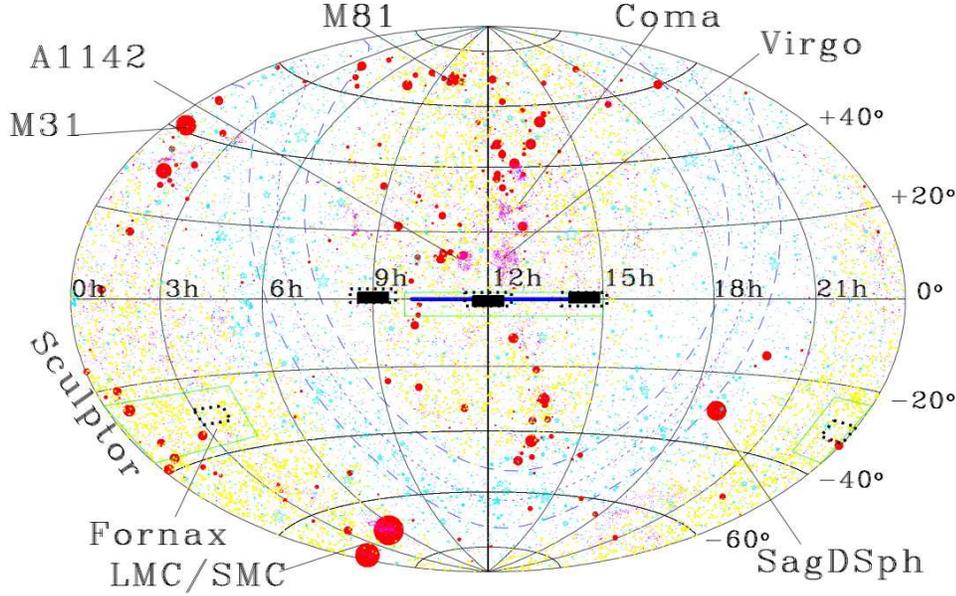} 
 \caption{(black rectangles) Survey regions of first year data obtained in 21
 clear nights as compared to the 2dFGRS (green outline) and MGC (blue
 band) surveys. Also shown are galaxies within 10Mpc (red dots), abell
 clusters (yellow), bright stars (cyan) and the NGC catalogue
 (magenta). The dashed regions show the final GAMA survey extent.}
   \label{fig1}
   \label{fig1}
\end{center}
\end{figure}

\begin{figure}[t]
\begin{center}
 \includegraphics[width=5.0in]{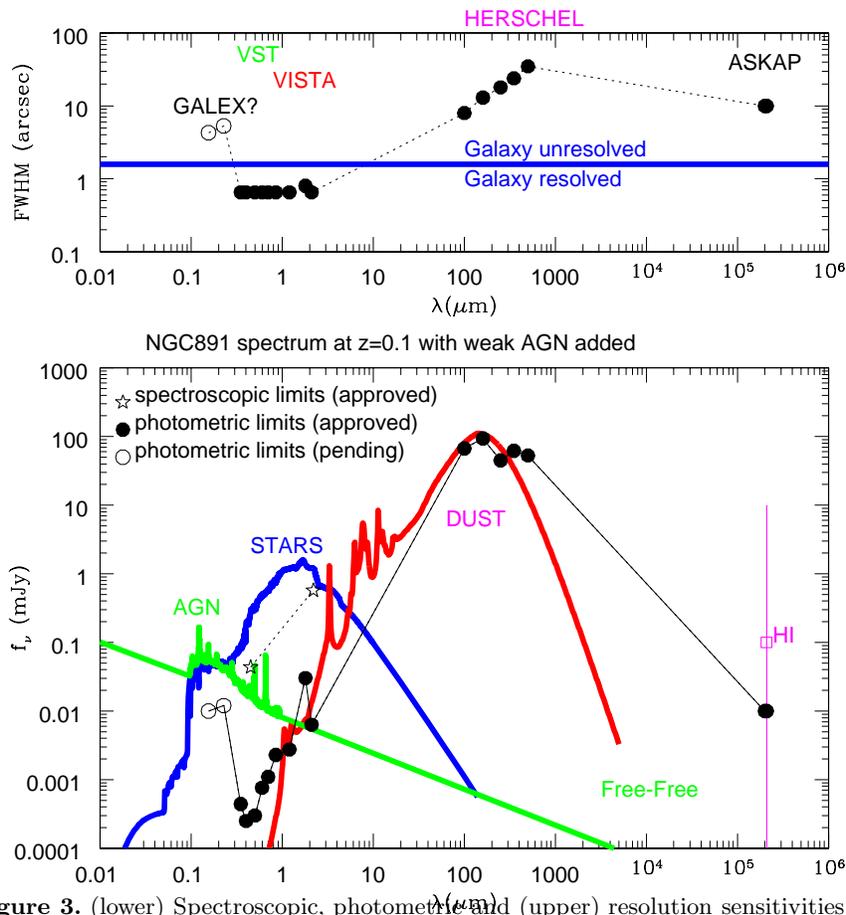} 
\vspace*{-1.0 cm}
 \caption{(lower) Spectroscopic, photometric and (upper) resolution
 sensitivities in mJsy and arcseconds respectively. Overlaid are the
 typical AGN (green), stellar (blue) and dust emissions (red; based on
 NGC891 at $z\approx0.1$ with weak AGN added).}
\end{center}
\end{figure}

\begin{figure}[t]
\begin{center}
 \includegraphics[width=5.0in]{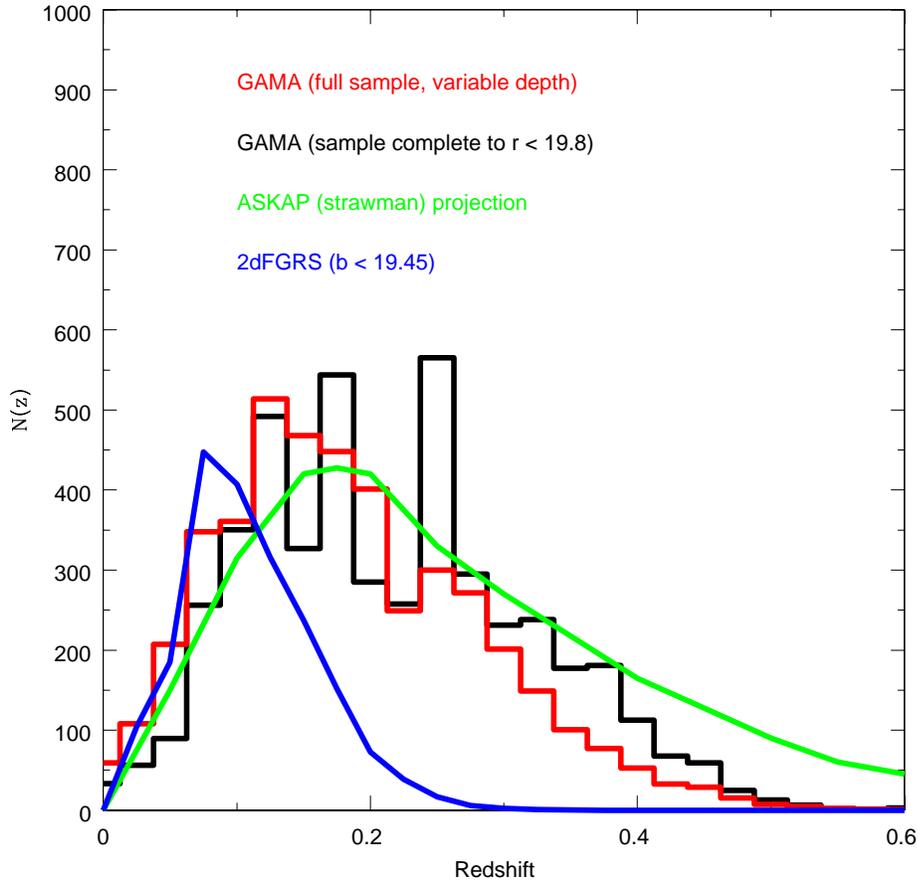} 
 \caption{The GAMA n(z) distribution compared to that expected by the ASKAP 1 year deep stare, aribtrarily normalised to give a comparable height peak.}
   \label{fig1}
\end{center}
\end{figure}

\begin{figure}[t]
\begin{center}
 \includegraphics[width=5.0in]{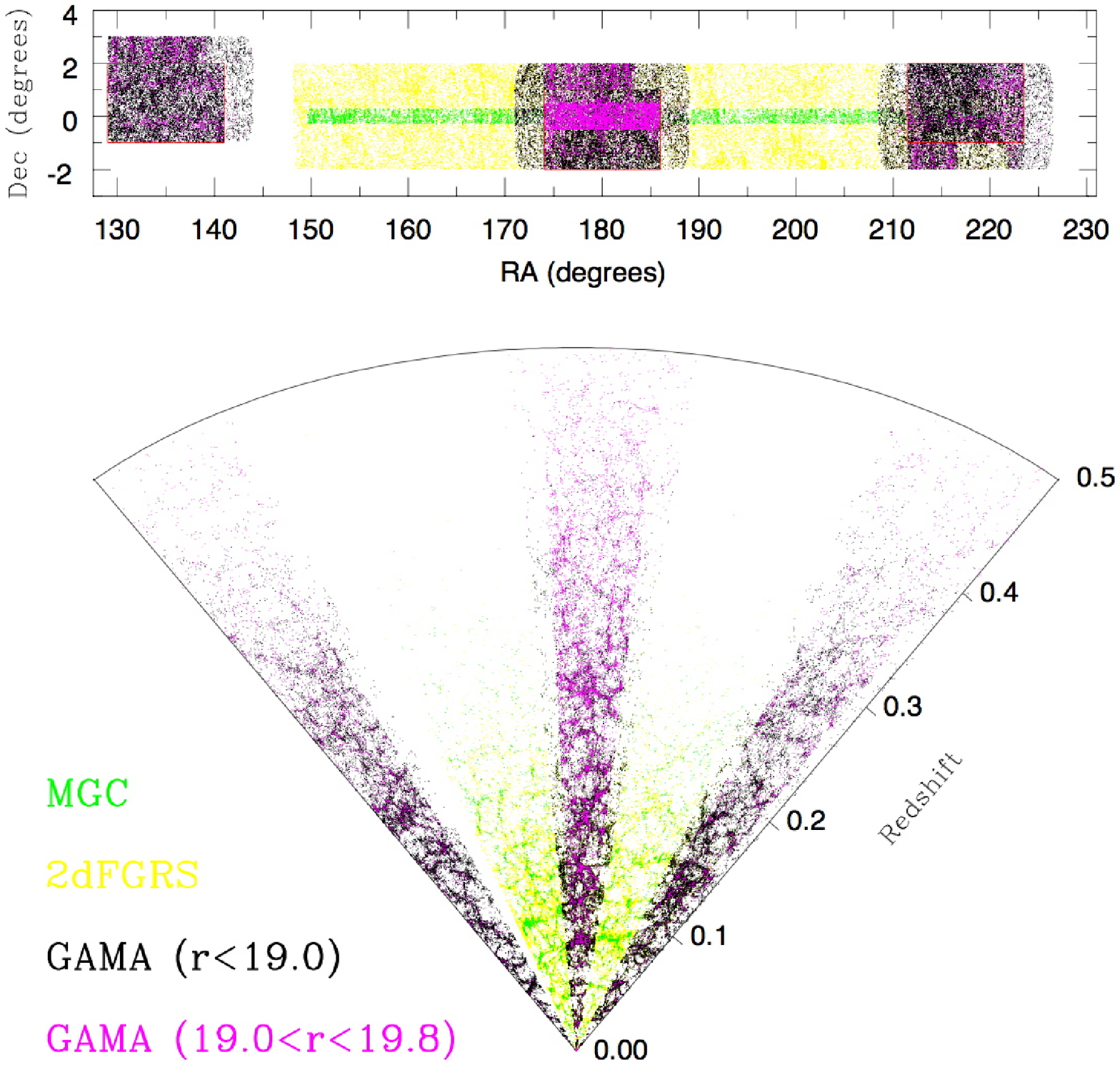} 
 \caption{(upper panel) The GAMA sky coverage and (lower panel) cone plot from the first year of GAMA observations at the AAT.}
   \label{fig1}
\end{center}
\end{figure}

\section{GAMA Motivation}
Galaxy And Mass Assembly (GAMA) is a major expansion of the Millennium
Galaxy Catalogue (MGC) survey (Liske et al 2003; Allen et al 2005;
Driver et al 2005) and a natural extension of the extremely productive
nearby ``Legacy'' surveys (e.g., SDSS, 2MASS, HIPASS etc).  In
comparison to the superb SDSS survey GAMA will only sample 250 sq
degrees of sky but will extend to significantly fainter spectroscopic
limits (12$\times$ the redshift density of SDSS main, 5$\times$ stripe
82), to higher spatial ($0.6''$ FWHM) and spectral (3---7\AA)
resolutions, as well as moving to a far broader wavelength coverage
(UV to Radio). GAMA has come about by parallel technological
developments leading to a suite of new facilities whose survey
sensitivities, resolutions, and capabilities are reasonably well
matched. Until now the study of galaxies has generally been restricted
to either large samples of limited wavelength data or multi-wavelength
studies of small (and often biased) samples. However galaxy systems
are extremely complex and diverse, exhibiting strong environmental and
mass dependencies and containing distinct but interlinked components
(AGN, nucleus, bulge, pseudo-bulge, bar, disc etc) and constituents
(SMBH, plasma, stars, gas, dust etc). It then follows that a clear
understanding of galaxy formation and evolution may only come about
via the construction of a comprehensive survey which simultaneously
samples all of these facets.  The GAMA team aims to provide this
data. In addition to the provision of a generic galaxy database, the
GAMA project also includes a number of more focussed science goals, in
particular:

\begin{description}
\item{1. } Measurement of the Halo Mass Function via virialised group velocity dispersions to directly test the {\it numerical} prediction from CDM (and WDM) simulations.
\item{2. } Measurement of the dynamic, baryonic, HI and stellar mass functions to LMC masses versus redshift, environment, type, and component (as well as higher order relations, e.g., mass-spin [$M-\lambda$]).
\item{3. } Measurement of the recent merger rates and star formation rates versus type, mass and environment over a 3---4 Gyr baseline.
\end{description}

\section{Facilities Contributing to GAMA}
Fig.~1 shows the facilities currently contributing to the GAMA project
along with the approximate time allocations within the GAMA sky
regions (see Fig.~2 and Tables~1~\&~2).  The expected source
resolution and detection sensitivities (5$\sigma$ point source) are
shown in Fig.~3 (upper and lower) in arcseconds and
milliJanskys. Overlaid on the lower panel is the modelled NGC891
spectra (Popescu et al., 2000) with a weak AGN added and transposed to $z
\approx 0.1$. The UKIRT data is provided courtesy of the UKIDSS LAS
Public Survey while the VST and VISTA data are provided via the KIDS
and VIKING ESO Public Surveys (whose teams include GAMA members).  The
Herschel data is provided as part of the broader Herschel-ATLAS survey
and a proposal is currently pending to complete GALEX medium depth
observations of the GAMA regions ($\sim$50\% already covered with
MIS). A major advancement over previous surveys will be the inclusion
of radio data via ASKAP (Australian Square Kilometer Array Pathfinder,
see Johnston et al. 2007) which should allow HI mass, dynamical mass
and continuum measurements for all GAMA galaxies with high or normal
neutral gas content. The deep ASKAP pointing (a single $6^o \times
6^o$ field of 1 year integration) is predicted to have an n(z)
distribution comparable to that derived for GAMA (see Fig.~4). While
the exact location of the ASKAP deep pointing has not been finalised
it is highly likely, given the similarity in the n(z) distributions
that one of the GAMA fields will be adopted (nominally the 12hr
field).

\begin{table}[h]
\caption{The current extent of the GAMA survey showing the year 1
regions and two possible expansion options currently under
consideration.}
{
\begin{tabular}{|c|c|c|c|c|c|c|} \hline
GAMA & \multicolumn{2}{c|}{Year 1 Regions} &\multicolumn{2}{c|}{Extension 1} & \multicolumn{2}{c|}{Extension 2} \\ \cline{2-7}
Field ID & RA(deg) & $\delta$(deg) & RA(deg) & $\delta$(deg) & RA(deg) & $\delta$(deg) \\ \hline
G09 & {\bf 129.0 --- 141.0} & {\bf -1 --- +2} & 129.0 --- 141.0 & -1 --- +3 & 129.0 --- 141.0 & -3 --- +3 \\ 
G12 & {\bf 174.0 --- 186.0} & {\bf -2 --- +1} & 174.0 --- 186.0 & -2 --- +2 & 129.0 --- 186.0 & -3 --- +3 \\ 
G14 & {\bf 211.5 --- 223.5} & {\bf -1 --- +2} & 211.5 --- 223.5 & -2 --- +2 & 211.5 --- 223.5 & -3 --- +3 \\ 
G03 & --- & --- & 45.0 --- 57.0 & -28 --- -31 & --- &  --- \\ 
G22 & --- & --- & 348.0 --- 360.0 & -28 --- -31 & --- &  --- \\ \hline 
\end{tabular}
} 
\end{table}

\begin{table}[h]
\caption{Time allocations, resolutions and sensitivities of the
facilities contributing to the GAMA survey. Limits are for 5$\sigma$
detections in AB mag or mJsky ($1J{\rm sky}=3631 \times
10^{-0.4m_{AB}}$).}
{\scriptsize
\begin{tabular}{|c|c|c|c|c|c|c|c|c|c|} \hline
Window & Facility & Collab.(Time) & \multicolumn{5}{c|}{Detection limits} & Resol. & GAMA Fields \\ \hline
UV & GALEX  & MIS +    & \multicolumn{2}{c}{FUV} & \multicolumn{2}{c}{NUV} &  & FUV/NUV & G09,G12,G15 \\ \cline{4-9}
   &        & 25hrs pend. & \multicolumn{2}{c }{23.0} & \multicolumn{2}{c}{23.0} &  &  4---5$''$ & \\ \hline
Opt & SDSS & DR6  & u & g & r & i & z & u-z & G09,G12,G15 \\ \cline{4-9}
    &      &      & 22.0 & 22.0 & 22.2  & 21.3  & 20.5  &  1.0$''$-2.5$''$ & \\ \cline{2-10}
    & VST & KIDS (70n) & 24.8 & 25.4 & 25.2 & 24.2 & --- & 0.6$''$ --- 1.0$''$ & All \\ \hline
Near-IR & UKIRT  & LAS (10n)  & Z & Y & J & H & K &  Z-K & G09,G12,G15 \\ \cline{4-9}
        &  &   & ---  & 20.9 & 20.6 & 20.3 & 20.1 & 0.6$''$-1.0$''$ & \\ \cline{2-10}
        & VISTA & VIKING (40n) & 23.1 & 22.3 & 22.1 & 21.5 & 21.1 & 0.6$''$---0.8$''$ & All \\ \hline
Far-IR &  Herschel  & ATLAS (200hrs) & 110$\mu$m & 170$\mu$m & 250$\mu$m & 350$\mu$m & 500$\mu$m & 110---500$\mu$m & \\ \cline{4-9}
       & & & 67mJ & 94mJ & 45mJ & 62mJ & 53mJ & 8---35$''$ & All \\ \hline
Radio &  ASKAP     & DEEP (1 year)       & \multicolumn{5}{c|}{0.7-1.8 GHz} & 0.7-1.8GHz & G12? \\ \cline{4-9} 
      &  & & \multicolumn{5}{c|}{10$\mu$J} & $\sim$10$''$ & \\ \hline
\end{tabular}
}
\end{table}


\section{First Light}
The survey commenced March 2008 with 50k spectra obtained in 21 clear
nights using the Anglo Australian Observatory's new multi-fibre-fed
bench-mounted dual-beam spectroscopic system (AA$\Omega$). Fig.~5
shows the areas of sky surveyed (upper) and the resulting cone
plot (lower). This includes the existing 25k redshifts within these
regions from the MGC, SDSS and 2dFGRS. AA$\Omega$ represents an
upgrade of the pre-existing 2dF system using the same fibre
positioner/tumbler but replacing the two telescope mounted
spectrographs with a single bench-mounted, double-beam spectrograph
(see Sharp et al. 2006 or the AAO website). The facility can be used
for both multi-fibre and integral field spectroscopy and in
multi-fibre mode is capable of obtaining 350---400 spectra in a single
2$^o$ diameter field. During an 8hr observation period the system is
capable of obtaining $\sim$3000 spectra. Data are reduced in real-time
and redshifts also obtained in real-time via cross-correlation with a
template library. All data are later re-reduced and processed with
GANDALF (see Schawinski et al 2007) to obtain line indices and velocity dispersion
measurements. The GAMA survey at the AAT uses the 580V and 385R
gratings yielding a resolution of 1300 or 3---7\AA FWHM.

\section{Summary}
The GAMA project has commenced with data flows imminent from a number
of international facilities. The survey will allow for a comprehensive
study of structure on 1kpc to 1Mpc scales as well as the subdivision
of the galaxy population into its distinct components and constituents.
Progress and data releases (1st data release forecast for Dec 2009) can be monitored
via the GAMA website: http://www.eso.org/$\sim$jliske/gama/ and anyone
interested in further details should contact Simon Driver at
spd3@st-and.ac.uk.

The GAMA team acknowledges all staff of the Anglo Australian
Observatory for the provision of the superb AA$\Omega$ facility and
the continued smooth running of the Anglo Australian Telescope. SPD
wishes to thank the IAU254 organisers for a most enjoyable meeting.

~

\noindent
Allen P.D., et al., 2005, MNRAS, 371, 2
$\bullet$  Driver S.P., et al., 2005, MNRAS, 360, 81
$\bullet$  Driver S.P., et al., 2008, ApJL, 678, 101 
$\bullet$  Johnston S., et al., 2007, PASA, 24, 174
$\bullet$  Liske J., et al., 2003, MNRAS, 344, 307
$\bullet$  Popescu C.C., et al 2000, A\&A, 362, 138 
$\bullet$  Sharp R., et al., 2006, SPIE, 6269, 14
$\bullet$  Schawinski et al., 2007, MNRAS, 382, 1415

\end{document}